\documentclass[11pt]{revtex4}
\usepackage{amsmath,mathrsfs}
\usepackage{graphicx,epsfig}
\usepackage{latexsym,amssymb}
\usepackage{epsf}
\usepackage{ifpdf,natbib,lineno}
\begin{document}
\title{Bertrand Spaces and Projected Closed Orbits in General Relativity}
\author{M. Rahimkhanli$^{1}$\\and\\N. Riazi$^{2}$\thanks{Electronic address: n\_riazi@sbu.ac.ir}\\
$^{1}$\emph{Physics Department and Biruni Observatory,}\\ \emph{Shiraz University, Shiraz 71454, Iran}\\
$^{2}$\emph{Department of Physics, Shahid Beheshti University,}\\
\emph{Evin, Tehran 19839, Iran}}
\begin{abstract}
In the present work, metrics which  lead to projected closed orbits
are found by comparing the relativistic differential equation of orbits
 with the corresponding classical differential equation. Physical and
 geometrical properties of these peculiar spacetimes are derived and discussed.
 It is also shown that some of these spacetimes belong to the broader class of the Bertrand spacetimes.
\end{abstract}
\maketitle
\section{Introduction}
\indent In classical mechanics, the existence of closed orbits is
very important, since it guides us toward discovering further
symmetries of the dynamical system. In general, closed-ness of
orbits for a specific angular momentum depends on energy. In order
to check the long term behavior, the stability of orbits should
also be considered \cite{Goldstein2002,Fowles1962,Marion1995}.
Classical Bertrand's theorem states that only two types of
potentials produce stable, closed orbits: the Kepler potential and
the harmonic oscillator potential
\cite{Goldstein2002,Bertrand1873}.

\indent Classical laws of gravitation in Newtonian mechanics, gained many successes in describing the gravitational phenomena, but they couldn't explain some of the observations such as the precession of the planetary orbits. Hence, a successful understanding of the gravitation required a new approach. This approach was introduced in the form of the general theory of relativity by Albert Einstien in 1915. Precession of the orbits is explained by this theory, that of Mercury around the Sun being the most famous example. One of the most important solutions of the Einstein's field equations, is the Schwarzschild solution. Within this metric, orbits are not closed. This example shows that the general relativistic bound orbits in vacuum are not generally closed, but rather they form precessing ellipses, as projected onto a spacelike hyper surface \cite{Weinberg1972,Inverno1992,Stephani2004}.

\indent In 1992, Perlick \cite{Perlick1992} showed that Bertrand's theorem can be reformulated in general relativity. This proposition is as the general relativistic analogue of the classical Bertrand theorem and the resulting spacetimes are known as Bertrand spacetimes. Bertrand spacetimes are interesting in their own right at least because of their mathematical properties. From the point of view of manifold theory, closed geodesics have long played a preponderant role in Riemannian geometry. A somewhat similar question, that of characterizing all Riemannian manifolds whose geodesics are all closed, is still wide open \cite{Klingenberg1983,Besse1987}. Under certain physical assumptions, the dark matter distribution of some low surface brightness galaxies can be described in terms of a particular class of the Bertrand spacetimes \cite{Dipanjan2013}.

\indent In this work, we first review the problem of closed orbits
in classical mechanics. The equation of orbit in general
relativity is presented in section 3. Projected closed orbits in
GR are then found by comparing the relativistic equation of orbit
with the corresponding classical equation (section 4). In section
5, the physical and geometrical properties of the resulting
spacetimes are derived and discussed. We introduce the Bertrand
spacetimes and discuss their relevance to some of the spacetimes
found in the present work in section 6. The final section contains
our concluding remarks.

\section{Equations of orbit in classical mechanics}

\indent Closed orbits appear in central forces and potentials in dynamics. We restrict ourselves to conservative central forces, where the potential $V\left(r\right)$ is a function of $r$ only. Since the potential energy (and thus the Hamiltonian) involves only the radial distance, the problem has spherical symmetry and the total angular momentum vector  ${\mathbf L}$ , is conserved:
\begin{equation} \label{1}
{\it {\mathbf L}}={\it {\mathbf r}}\times {\it {\mathbf p}} ,
\end{equation}
\begin{equation} \label{2}
|{\it {\mathbf L}}|=mr^{2} \dot{\varphi }\equiv L=constant,
\end{equation}
where ${\mathbf r}$ is the radial vector,  ${\mathbf p}$ is the linear momentum, $m$ is the test mass, $\varphi $ is angular coordinate in the plane of orbit, and $L$ is the magnitude of angular momentum. Since the force is conservative, on the basis of the general energy conservation theorem, the total energy$\ E$, is a constant of motion:
\begin{equation} \label{3}
E=\frac{1}{2} m(\dot{r}^{2} +r^{2} \dot{\varphi }^{2} )+V(r)=constant.
\end{equation}
\indent Consider a test particle of unit mass $(m=1)$ moving in a conservative, spherically symmetric, attractive potential within the framework of classical mechanics. The total energy and angular momentum of this particle are constants of motion.\textbf{ }The resulting orbit is then confined to a plane, which we conveniently identify with the (x,y)-plane (i.e. $\theta =\frac{\pi }{2}$). After a little algebra, we arrive at the following equation of orbit:\textbf{}
\begin{equation} \label{4}
(\frac{du}{d\varphi } )^{2} +u^{2} +\frac{2}{L^{2} } \; V-\frac{2E}{L^{2} } =0,
\end{equation}
in which $L$ and $E$ are constants and $u=\frac{1}{r}$ . For the
Kepler potential, $\ V=-\frac{1}{r}=-u$ , (we assume the potential
coefficients $k = 1$), the equation of orbit \ref{4} becomes
\begin{equation} \label{5}
(\frac{du}{d\varphi } )^{2} +u^{2} -\frac{2}{L^{2} } \; u-\frac{2E}{L^{2} } =0.
\end{equation}
By taking the derivative of equation \ref{5}, a simpler, linear
equation results
\begin{equation} \label{6}
\frac{d^{2} u}{d\varphi ^{2} } +u=\frac{1}{L^{2} } \; ,
\end{equation}
which has the following closed orbit solution \cite{Goldstein2002}:
\begin{equation} \label{7}
u=\frac{1}{r} =\frac{1}{L^{2} } [1+\sqrt{1+2EL^{2} } cos(\varphi -\varphi _{0} )],
\end{equation}
in which ${\varphi }_o$ is the initial value of $\varphi $. The
orbits are ellipses with one focus located at the center of
potential $r=0$ (Kepler's first law). For the classical motion of
a unit mass test particle inside a 3D harmonic potential,
$V=\frac{1}{2}r^2=\frac{1}{2u^2}$, (we assume the potential
coefficients $k' = 1$), the equation of orbit \ref{4} becomes
\begin{equation} \label{8}
(\frac{du}{d\varphi } )^{2} +u^{2} +\frac{1}{L^{2} } \; \frac{1}{u^{2} } -\frac{2E}{L^{2} } =0.
\end{equation}
Equation \ref{8}, after taking a derivative becomes
\begin{equation} \label{9}
\frac{d^{2} u}{d\varphi ^{2} } +u=\frac{1}{L^{2} } \frac{1}{u^{3} } ,
\end{equation}
which has the following closed orbit solution:
\begin{equation} \label{10}
u^{2} =\frac{1}{r^{2} } =a^{2} {\rm cos}^{2} \varphi +b^{2} {\rm sin}^{2} \varphi ,
\end{equation}
in which $a^{-1}$ and $b^{-1}$ are semi-major and semi-minor axes of the ellipse, respectively. Here, like in the Kepler problem,\textbf{ }the orbits are also ellipses, but this time with center at $r=0$. According to the Bertrand's theorem, the Kepler and harmonic potentials are the only attractive potentials in classical mechanics which lead to closed bound orbits \cite{Goldstein2002,Bertrand1873}. These potentials and only these, could possibly produce closed orbits for any arbitrary combination of $L$ and $E\ (E<0)$ \cite{Goldstein2002,Bertrand1873,Arnold1989}. The proof of Bertrand's theorem is not difficult, and has actually been included (with various levels of rigor) in several textbooks and papers (for example see \cite{Bertrand1873} or \cite{Santos2007}).

\indent The closed-ness of orbits in classical mechanics signals the existence of extra constants of motion besides the total energy and angular momentum. It can be shown that for the Kepler problem, the  following vector (known as the Laplace-Runge-Lenz vector) is a constant of motion \cite{Goldstein2002}:
\begin{equation} \label{11}
{\it {\mathbf A}}={\it {\mathbf p}}\times {\it {\mathbf L}}-\widehat{{\it {\mathbf r}}},
\end{equation}
where ${\widehat{{\it {\mathbf r}}}=\frac{\mathbf r}{r}}$. This vector has become known amongst physicists as the Runge-Lenz vector, but priority belongs to Laplace \cite{Goldstein2002}.\textbf{ }The Laplace-Runge-Lenz vector for the 3D harmonic potential reads \cite{Fradkin1976}\textbf{     }
\begin{equation} \label{12}
{\it {\mathbf A}}=f\left({\mathbf r},L,E,\omega \right)[{\it {\mathbf p}}\times {\it {\mathbf L}}]-g\left({\mathbf r},L,E,\omega \right)\widehat{{\it {\mathbf r}}},
\end{equation}
in which $\omega $ is the frequency of the oscillator; $f$ and ${\rm \textit{g}}$ are certain functions of ${\mathbf r}$ and constants of motion.

\section{Equations of orbit in general relativity}
\indent Let us now turn to general relativity. In GR, equations of motion of a freely falling particle, known as geodesic equations, are \cite{Weinberg1972,Inverno1992,Stephani2004}:
\begin{equation} \label{13}
\frac{d^{2} x^{\mu } }{d\lambda ^{2} } +{\rm \Gamma }_{\; \; \nu \kappa }^{\mu } \frac{dx^{\nu } }{d\lambda } \frac{dx^{\kappa } }{d\lambda } =0,
\end{equation}
where ${\Gamma }^{\mu }_{\ \ \nu \kappa }$ is the affine connection and $\lambda $ is the affine parameter. Consider orbits in a static, spherically symmetric spacetime \cite{Weinberg1972}:
\begin{equation} \label{14}
ds^{2} =-B(r)dt^{2} +A(r)dr^{2} +r^{2} d\theta ^{2} +r^{2} sin^{2} \theta d\varphi ^{2} .
\end{equation}
By using this metric, the geodesic equations \ref{13} take the
following forms \cite{Weinberg1972}: \textbf{}
\begin{equation} \label{15}
\frac{d^{2} r}{d\lambda ^{2} } +\frac{A'}{\; 2A} (\frac{dr}{d\lambda } )^{2} -\frac{r}{\; A} (\frac{d\theta }{d\lambda } )^{2} -\frac{rsin^{2} \theta }{\; A} (\frac{d\varphi }{d\lambda } )^{2} +\frac{B'}{\; 2A} (\frac{dt}{d\lambda } )^{2} =0,
\end{equation}
\begin{equation} \label{16}
\frac{d^{2} \theta }{d\lambda ^{2} } +\; \frac{2}{\; r} \frac{dr}{d\lambda } \frac{d\theta }{d\lambda } -\; sin\theta \; cos\theta (\frac{d\varphi }{d\lambda } )^{2} =0,
\end{equation}
\begin{equation} \label{17}
\frac{d^{2} \varphi }{d\lambda ^{2} } +\frac{2}{\; r} \frac{dr}{d\lambda } \frac{d\varphi }{d\lambda } +2cot\theta \frac{d\theta }{d\lambda } \frac{d\varphi }{d\lambda } =0,
\end{equation}
\begin{equation} \label{18}
\frac{d^{2} t}{d\lambda ^{2} } +\frac{B'}{B} \frac{dr}{d\lambda } \frac{dt}{d\lambda } =0,
\end{equation}
where prime denotes $d/dr$. Since the field is isotropic, we may consider the orbit of our particle to be confined to the equatorial plane, that is $\theta =\frac{\pi }{2}(=const.)$ \cite{Weinberg1972,Inverno1992}. Thus, the geodesic equations become:
\begin{equation} \label{19}
\frac{d^{2} r}{d\lambda ^{2} } +\frac{A'}{\; 2A} (\frac{dr}{d\lambda } )^{2} -\frac{r}{\; A} (\frac{d\varphi }{d\lambda } )^{2} +\frac{B'}{\; 2A} (\frac{dt}{d\lambda } )^{2} =0,
\end{equation}
\begin{equation} \label{20}
\frac{d^{2} \varphi }{d\lambda ^{2} } +\frac{2}{\; r} \frac{dr}{d\lambda } \frac{d\varphi }{d\lambda } =0,
\end{equation}
\begin{equation} \label{21}
\frac{d^{2} t}{d\lambda ^{2} } +\frac{B'}{B} \frac{dr}{d\lambda } \frac{dt}{d\lambda } =0.
\end{equation}
Now, we are going to obtain the relativistic equations of the
orbit. We can rewrite relations \ref{20} and \ref{21} in the
folowing forms:
\begin{equation} \label{22}
\frac{d^{2} \varphi }{d\lambda ^{2} } +\frac{2}{\; r} \frac{dr}{d\lambda } \frac{d\varphi }{d\lambda } =\frac{1}{\; r^{2} } \frac{d}{d\lambda } \left(r^{2} \frac{d\varphi }{d\lambda } \right)=0,
\end{equation}
\begin{equation} \label{23}
\frac{d^{2} t}{d\lambda ^{2} } +\frac{B'}{B} \frac{dr}{d\lambda } \frac{dt}{d\lambda } =\frac{1}{\; B} \frac{d}{d\lambda } \left(B\frac{dt}{d\lambda } \right)=0.
\end{equation}
From above, there result two constants of motion
\begin{equation} \label{24}
r^{2} \frac{d\varphi }{d\lambda } \equiv J={\rm constant},
\end{equation}
\begin{equation} \label{25}
B\frac{dt}{d\lambda } ={\rm constant}.
\end{equation}
Since $\lambda $ is an arbitrary affine parameter, we can set the constant term $B\frac{dt}{d\lambda }$ equal to 1:
\begin{equation} \label{26}
B\frac{dt}{d\lambda } =1.
\end{equation}
From \ref{24} we see that \textit{J} (that is angular momentum per
unit mass of particle) is a constant of motion. Using relations
\ref{24} and \ref{26} in the geodesic equation \ref{19}, and also
the following identities:
\begin{equation} \label{27}
\frac{d}{d\lambda } \left[A\left(\frac{dr}{d\lambda } \right)^{2} \right]=2A\frac{dr}{d\lambda } \frac{d^{2} r}{d\lambda ^{2} } +A'\left(\frac{dr}{d\lambda } \right)^{3} ,
\end{equation}
\begin{equation} \label{28}
\frac{d}{d\lambda } \left[\frac{J^{2} }{\; r^{2} } -\frac{1}{\; B} \right]=-\frac{\; 2J^{2} }{\; r^{3} } \frac{dr}{d\lambda } +\frac{B'}{\; B^{2} } \frac{dr}{d\lambda }  ,
\end{equation}
we arrive at
\begin{equation} \label{29}
\frac{d}{d\lambda } \left[A\left(\frac{dr}{d\lambda } \right)^{2} +\frac{J^{2} }{\; r^{2} } -\frac{1}{\; B} \right]=0.
\end{equation}
Equation \ref{29}, shows that the quantity
$A{\left(\frac{dr}{d\lambda }\right)}^2+\frac{J^2}{\
r^2}-\frac{{\rm 1}}{\ B}$  is another constant of motion:
\begin{equation} \label{30}
A\left(\frac{dr}{d\lambda } \right)^{2} +\frac{J^{2} }{\; r^{2} } -\frac{1}{\; B} \equiv -\varepsilon .
\end{equation}
From equation \ref{30}, one easily obtains the following equation
for the orbit in general relativity:
\begin{equation} \label{31}
(\frac{du}{d\varphi } )^{2} +\frac{1}{\; A} u^{2} -\frac{1}{\; J^{2} AB} +\frac{\varepsilon }{\; J^{2} A} =0,
\end{equation}
in which  \textit{J}   and $\varepsilon $ are constants, $u=\frac{1}{r}$, and  \textit{A}  and  \textit{B}  are functions
of $r$ only.
\indent The Schwarzschild solution in the Boyer-Lindquist\textbf{ }form ($G=c=1)$ is given by \cite{Inverno1992}:
\begin{equation} \label{32}
ds^{2} =-(1-\frac{2M}{\; r} )dt^{2} +\; (1-\frac{2M}{\; r} )^{-1{\rm \; }} dr^{2} +r^{2} d\theta ^{2} +r^{2} sin^{2} \theta d\varphi ^{2} .
\end{equation}
For this metric (with unit mass$\ M=1$), the functions $\ A\left(r\right)$  and $B\left(r\right)$ read:
\begin{equation} \label{33}
A(r)=\; (1-2u)^{-1{\rm \; }} , \ \ \    B(r)=(1-2u)^{{\rm \; }} =A^{-1} ,
\end{equation}
and the equation of orbit \ref{31}, takes the following form:
\begin{equation} \label{34}
\left(\frac{du}{d\varphi } \right)^{2} -2u^{3} +u^{2} -\frac{2\varepsilon }{\; J^{2} } u+\frac{1}{\; J^{2} } (1+\varepsilon )=0.
\end{equation}
It can be shown that this equation has the following approximate solution \cite{Inverno1992}:
\begin{equation} \label{35}
u\simeq \frac{\varepsilon }{\; J^{2} } [1+e\; {\rm cos}((1-\delta )\varphi )],
\end{equation}
where $e$ is eccentricity of the orbit, and $\delta =\frac{2\varepsilon }{\ J^2}$ is constant. This result shows the famous precession of the perihelion, which implies that general relativistic bound orbits in vacuum are not generally closed, but rather they form precessing ellipses, that of Mercury around the Sun being the most famous example \cite{Weinberg1972,Inverno1992,Stephani2004}.

\section{Projected closed orbits in GR}

\indent Now we have both classical \ref{4} and relativistic
\ref{31} differential equations of orbit at hand. We know that the
classical equation, with the Kepler potential and the harmonic
potential, lead to closed orbits. We are going to find metrics
which lead to closed orbits within GR, by comparing the
relativistic differential equation of orbits with the
corresponding classical differential equation. Obviously, we have
to go beyond the Schwarzschild vacuum solution for this purpose.
We can compare the GR \ref{31} and Newtonian \ref{4} equations in
several different ways. One way is that we set the second term of
the equation \ref{31} equal to the second term of the equation
\ref{4}:
\begin{equation} \label{36}
\frac{1}{\; A} u^{2} =u^{2}
\end{equation}
or equivalently:
\begin{equation} \label{37}
A(r)=1.
\end{equation}
Now we obtain $B\left(r\right)$, by setting $A={\rm 1}$ in the
relativistic differential equation \ref{31}, and then setting the
resulting equation equal to classical differential equation
\ref{4}. This leads to
\begin{equation} \label{38}
B(r)=\frac{(-\frac{L^{2} }{\; 2J^{2} } )}{V(r)-[E+\frac{L^{2} }{\; 2J^{2} } \varepsilon ]} .
\end{equation}
In other words, we demand the GR orbit equation \ref{31} to take
apparently the form of the classical orbit equation \ref{4}. Now,
if we insert the Kepler potential, for $V\left(r\right)$ in
\ref{38}, then we turn the relativistic differential equation of
orbits to the differential equation of orbits which have exactly
the same form as the classical Kepler orbits (i.e. ellipses with
one focus located at the center). By setting $V(r){\rm
=}-\frac{1}{r}$  in \ref{38}, $B\left(r\right)$ takes the
following form:
\begin{equation} \label{39}
B(r)=C\; \frac{r}{r+C_{1} } ,
\end{equation}
in which $C={(\frac{2J^2}{\ L^2}E+\varepsilon )}^{-1}$ and
$C_1={(E+\frac{L^2}{\ 2J^2}\varepsilon )}^{-1}$ are constants.
From \ref{37} and \ref{39}, the metric \ref{14} reads
\begin{equation} \label{40}
ds^{2} =-\frac{r}{r+C_{1} } dt^{2} +dr^{2} +r^{2} d\theta ^{2} +r^{2} sin^{2} \theta d\varphi ^{2} ,
\end{equation}
where the constant \textit{C} is absorbed into the definition of
$t$. Since the metric \ref{40} leads to exactly the same equation
of the  orbit as the classical Kepler orbits, we have the
elliptical orbits. Similarly, if we set the harmonic potential,
for $V\left(r\right)$ in \ref{38}, then we arrive from the
relativistic differential equation of orbits to one which is
exactly the same differential equation of classical  harmonic
orbits (i.e. ellipses centered at  $r{\rm =0}$). By setting
$V(r){\rm =}\frac{1}{2}r^2$  in \ref{38}, $B\left(r\right)$ takes
the following form:          \textit{    }\textbf{  }
\begin{equation} \label{41}
B(r)=C\; \frac{1}{1+C_{2} r^{2} } ,
\end{equation}
in which  $C_2$$=-{(2E+\frac{L^2}{\ J^2}\varepsilon )}^{-1}$ is
constant. From \ref{37} and \ref{41}, the metric \ref{14} reads
\begin{equation} \label{42}
ds^{2} =-\frac{1}{1+C_{2} r^{2} } dt^{2} +dr^{2} +r^{2} d\theta ^{2} +r^{2} sin^{2} \theta d\varphi ^{2} ,
\end{equation}
where again the constant \textit{C} is absorbed into the definition of $t$. Since the metric \ref{42} leads to exactly the same equation of orbit as the classical harmonic orbits, we have the elliptical orbits in this spacetime too.\\
\indent We pointed out that we can compare the relativistic
differential equation of orbit \ref{31}, with its classical
counterpart \ref{4}, in several different ways. Until now, we
derived two of the possible options. As one can see from equations
\ref{31} and \ref{4}, there are three terms in the relativistic
differential equation which can be compared with the three terms
in the classical equation, i.e. we can arrange the correspondence
in 9 different ways. Since for each choice there are two possible
potentials (Kepler and harmonic), we have a total 18 possible
spacetimes using our form-compatibility method. Hitherto, we set
the second term of the equation \ref{31} equal to the second term
of the equation \ref{4}, and the functions \textit{A} and
\textit{B} read as \ref{37} and \ref{38}, then by using the Kepler
and harmonic potentials, metric \ref{40} and metric \ref{42} were
derived. Similary, we can set the second term of the equation
\ref{31} equal to the third or fourth term of the equation
\ref{4}, and so on. All 18 possible spacetimes derived in this
way, are listed in table \ref{tab:all spacetimes}.

\begin{table}
\caption{All possible spacetimes which obtained by using form-compatibility method.}
\label{tab:all spacetimes}
\begin{center}
\begin{tabular}{|p{0.4in}|p{0.8in}|p{0.6in}|p{0.4in}|p{0.8in}|p{0.6in}|} \hline
$metric$ & \textit{$B\left(r\right)$} & \textit{$A\left(r\right)$} & $metric$ & \textit{$B\left(r\right)$} & \textit{$A\left(r\right)$} \\ \hline
$1$ & $\frac{r}{r+C_{1} } $ & $1$ & $10$ & $\frac{r^{2} +C_{13} }{r^{4} +C_{11} r^{2} } $ & $\frac{r^{2} +C_{11} }{r^{2} +C_{13} } $ \\ \hline
$2$ & $\frac{1}{1+C_{2} r^{2} } $ & $1$ & $11$ & $\frac{r+C_{14} }{r^{2} +C_{11} } $ & $\frac{r^{2} +C_{11} }{r+C_{14} } $ \\ \hline
$3$ & $\frac{r^{3} }{r^{3} +C_{3} r^{2} +C_{4} } $ & $\frac{1}{r} $ & $12$ & $\frac{r^{4} +C_{15} }{r^{2} +C_{11} } $ & $\frac{r^{2} +C_{11} }{r^{4} +C_{15} } $ \\ \hline
$4$ & $\frac{r^{6} }{r^{6} +C_{5} r^{2} +C_{6} } $ & $\frac{1}{r^{4} } $ & $13$ & $\frac{r^{2} }{r^{4} +C_{10} r^{3} +C_{8} } $ & $r^{2} $ \\ \hline
$5$ & $\frac{r^{4} }{r^{4} +C_{7} r+C_{8} } $ & $\frac{1}{r^{2} } $ & $14$ & $\frac{r^{2} }{r^{6} +C_{12} r^{4} +C_{6} } $ & $r^{2} $ \\ \hline
$6$ & $\frac{r^{4} }{r^{4} +C_{9} } $ & $\frac{1}{r^{2} } $ & $15$ & $\frac{r^{2} }{r^{3} +C_{13} r+C_{7} } $ & $r$ \\ \hline
$7$ & $\frac{r^{4} +C_{10} r^{3} }{r^{2} +C_{11} } $ & $\frac{r^{2} +C_{11} }{r^{2} +C_{10} r} $ & $16$ & $\frac{r^{4} }{r^{2} +C_{16} } $ & $\frac{1}{r^{2} } $ \\ \hline
$8$ & $\frac{r^{6} +C_{12} r^{4} }{r^{2} +C_{11} } $ & $\frac{r^{2} +C_{11} }{r^{4} +C_{12} r^{2} } $ & $17$ & $\frac{r^{2} }{r+C_{17} } $ & $1$ \\ \hline
$9$ & $\frac{r^{3} +C_{13} r}{r^{2} +C_{11} } $ & $\frac{r^{2} +C_{11} }{r^{2} +C_{13} } $ & $18$ & $\frac{r^{2} }{r^{4} +C_{18} } $ & $1$ \\ \hline
\end{tabular}
\end{center}
\end{table}

\begin{table}
\caption{The constants which are used in the definition of the metrics.}
\label{tab:The constants}
\begin{center}
\begin{tabular}{|p{0.2in}|p{1in}|p{0.2in}|p{1in}|p{0.2in}|p{1in}|} \hline
$C_{1} $ & $(E+\frac{L^{2} }{\; 2J^{2} } \varepsilon )^{-1} $ & $C_{7} $ & $\frac{-J^{2} }{E\varepsilon } $ & $C_{13} $ & $\frac{-L^{2} }{\; 2E} $ \\ \hline
$C_{2} $ & $-(2E+\frac{L^{2} }{\; J^{2} } \varepsilon )^{-1} $ & $C_{8} $ & $\frac{L^{2} J^{2} }{2E\varepsilon } $ & $C_{14} $ & $\frac{-L^{2} }{\; 2} $ \\ \hline
$C_{3} $ & $\frac{-EJ^{2} }{\varepsilon } $ & $C_{9} $ & $(\frac{\; 2E\varepsilon }{L^{2} J^{2} } +\frac{1}{L^{2} } )^{-1} $ & $C_{15} $ & $L^{2} $ \\ \hline
$C_{4} $ & $\frac{L^{2} J^{2} }{2\varepsilon } $ & $C_{10} $ & $\frac{1}{\; E} $ & $C_{16} $ & $\frac{-L^{2} \varepsilon }{2E\varepsilon +J^{2} } $ \\ \hline
$C_{5} $ & $\frac{2EJ^{2} }{\varepsilon } $ & $C_{11} $ & $\frac{J^{2} }{\varepsilon } $ & $C_{17} $ & $\frac{-J^{2} E}{\varepsilon } -\frac{L^{2} }{2\; } $ \\ \hline
$C_{6} $ & $\frac{-L^{2} J^{2} }{\varepsilon } $ & $C_{12} $ & $-2E$ & $C_{18} $ & $\frac{2J^{2} E}{\varepsilon } +L^{2} $ \\ \hline
\end{tabular}
\end{center}
\end{table}

\section{Physical and geometrical structure of the spacetimes}

\indent In this section, we calculate the geometrical tensors of the spacetimes obtained in the previous section and discuss some of their properties.\\
\indent The metric \ref{40}, is asymptotically flat (Minkowski).
For this metric the components of the Einstein tensor read
\begin{equation} \label{43}
(G_{\; \; \nu }^{\mu } )=diag(\; \; 0, \ \frac{C_{1} }{r^{2} \left(r+C_{1} \right)} ,\ \frac{C_{1} \left(C_{1} -2r\right)}{4r^{2} \left(r+C_{1} \right)^{2} } , \ \; \frac{C_{1} \left(C_{1} -2r\right)}{4r^{2} \left(r+C_{1} \right)^{2} } ),
\end{equation}
and the Ricci scalar and the Kretschmann invariant read\textbf{}
\begin{equation} \label{44}
R=-\frac{3\; C_{1} ^{2} }{2r^{2} \left(r+C_{1} \right)^{2} } ,  \ \ \  K=\frac{3\; C_{1} ^{2} (8r^{2} +8C_{1} r+3\; C_{1} ^{2} )}{4r^{4} \left(r+C_{1} \right)^{4} } .
\end{equation}
Note that the Einstein tensor, the Ricci scalar and the
Kretschmann invariant are singular at $r{\rm =0\ }$ and $r{\rm
=-}C_1$, and vanish as $r\to \infty $. From the form of the
Einstein tensor through the Einstein equations $(G^{\mu }_{\ \ \nu
}=-8\pi  {\ T}^{\mu }_{\ \ \nu })$, we conclude that the following
density and pressure components are required to support the
metric:

\begin{equation} \label{45}
\rho =0,      p_{r} =-\frac{1}{\; 8\pi } G_{\; \; 1}^{1} =-\frac{C_{1} }{8\pi r^{2} \left(r+C_{1} \right)} ,     p_{t} =-\frac{1}{\; 8\pi } G_{\; \; 2}^{2} =-\; \frac{C_{1} \left(C_{1} -2r\right)}{32\pi r^{2} \left(r+C_{1} \right)^{2} } .
\end{equation}
It is seen that the energy density vanishes, and the weak energy condition \cite{Carroll2004,Poisson2004} is violated throughout the spacetime:
\begin{equation}\label{46}
\rho +p_{r} <\; 0
\end{equation}
(for  $C_1>0$)

\begin{equation}\label{47}
\rho +p_{t} <\; 0
\end{equation}
(for  $C_1<0$).\\
The line element \ref{40}, (if $\ C_1<0$), has a singularity \cite{Inverno1992,Misner1973,Wald1984} at $r_m=-C_1$. This singularity is a curvature singularity, since the Kretschmann invariant \ref{44}, is infinite at $r{\rm =}{\rm -}C_1$. For $r<r_m$, the metric signature becomes improper (i.e. it becomes non-Lorentzian or Euclidean).\\
Tolman mass-energy of a physical system, is given by the Tolman formula \cite{Florides2009}
\begin{equation} \label{48}
M=\int _{V}(-T_{\; \; \; 0}^{0} +T_{\; \; \; 1}^{1} +T_{\; \; \; 2}^{2} +T_{\; \; \; 3}^{3} )\sqrt{-g} dV ,
\end{equation}
in which $T^i_{\ \ j}$ is the energy-momentum-stress tensor of the system, and \textit{g} is determinant of the metric tensor. Thus, total mass of a spherically symmetric static spacetime is given by
\begin{equation} \label{49}
M=\int _{0}^{2\pi }\int _{0}^{\pi }\int _{0}^{\infty }(\; -T_{\; \; 0}^{0} +\; T_{\; \; 1}^{1} +\; T_{\; \; 2}^{2} +\; T_{\; \; 3}^{3} )\sqrt{-g} \; dr\; d\theta \; d\varphi    ,
\end{equation}
where for the spherically symmetric static spacetime \ref{14},
$\sqrt{-{\rm \textit{g}}}\ $ reads
\begin{equation} \label{50}
\sqrt{-{\rm \textit{g}}} =\sqrt{A(r)B(r)} ^{} r^{2} sin\theta .
\end{equation}
From \ref{49},\ref{45}, and \ref{40}, total mass of the spacetime
reads
\begin{equation}\label{51}
M=-\frac{1}{\; 2} C_{1}
\end{equation}
Note that for  $C_1{\rm <}0$, the total mass is positive.\\
\indent Metric \ref{42}, is asymptotically non-flat. For this
metric, the components of the Einstein tensor, the Ricci scalar
and the Kretschmann invariant read
\begin{equation} \label{52}
(G_{\; \; \nu }^{\mu } )=diag(0, \ -\frac{2C_{2} }{1+C_{2} r^{2} } , \ -\frac{C_{2} \left(2-C_{2} r^{2} \right)}{\left(1+C_{2} r^{2} \right)^{2} } , \ -\frac{C_{2} \left(2-C_{2} r^{2} \right)}{\left(1+C_{2} r^{2} \right)^{2} } ),
\end{equation}
\begin{equation} \label{53}
R=\frac{6C_{2} }{\left(1+C_{2} r^{2} \right)^{2} } , \ \ \
   K=\frac{12\; C_{2} ^{2} (1+2\; C_{2} ^{2} r^{4} )}{\left(1+C_{2} r^{2} \right)^{4} } .
\end{equation}
These geometrical quantities are regular everywhere if  $C_2>0$, and asymptotically tend to constants as  $r\to 0$  or  $r\to \infty $. The energy density and pressure components for this metric read
\begin{equation} \label{54}
\rho =0,     p_{r} =-\frac{1}{\; 8\pi } G_{\; \; 1}^{1} =\frac{C_{2} }{4\pi (1+C_{2} r^{2} )} ,     p_{t} =-\frac{1}{8\pi } G_{\; \; 2}^{2} =\; \frac{C_{2} \left(2-C_{2} r^{2} \right)}{8\pi \left(1+C_{2} r^{2} \right)^{2} } .
\end{equation}
We see that for  $r{\rm >}\sqrt{\frac{2}{\ C_2}}$  , WEC is violated throughout the spacetime:
\begin{equation} \label{55}
\rho +p_{t} <\; 0,
\end{equation}
while for  $r{\rm <}\sqrt{\frac{2}{\ C_2}}$ , although the energy density vanishes, but the pressure components satisfy the weak energy condition:
\begin{equation} \label{56}
\rho =\; 0 ,  \ \ \     \rho +p_{i} \ge \; 0.
\end{equation}
From \ref{54}, for  $C_2\ge 0$ we have
\begin{equation} \label{57}
\rho +p_{r} +p_{t} +p_{t} =\frac{3C_{2} }{4\pi \left(1+C_{2} r^{2} \right)^{2} } \ge 0.
\end{equation}
From \ref{57} and \ref{56}, we see that the strong energy condition \cite{Poisson2004,Wald1984} is satisfied for  $r<\sqrt{\frac{2}{\ C_2}}$ \ if $C_2{\rm >}0$. Thus, for this spacetime ($C_2>0$) there is a radius ($r_h=\sqrt{\frac{2}{\ C_2}}$) below which the WEC and SEC are satisfied.\\
If  $C_2<0$, the line element \ref{42} has a singularity at  $r_m{\rm =}\frac{1}{\sqrt{-C_2}}$. Since the Kretschmann invariant \ref{53}, is infinite at $r{\rm =}\frac{1}{\sqrt{-C_2}}$, this singularity is a curvature singularity.\\
From \ref{49},\ref{54}, and \ref{42}, total mass of the spacetime
reads
\begin{equation} \label{58}
M=\frac{1}{\sqrt{C_{2} } } .
\end{equation}
We see that for $C_2{\rm >}0$, the total mass is real and positive. Therefore this metric may be a perfect fluid solution of the Einstein's field equations.\\
\indent For the metric 3 to the metric 18 in table \ref{tab:all spacetimes}, components of the Einstein tensor and the Ricci scalar are given in table \ref{tab:Einstein tensor} and table \ref{tab:Ricci scalar} respectively. Some properties of these spacetimes, are given in table \ref{tab:properties}.\\

\begin{table}
\caption{The components of the Einstein tensor for all the 18 classes of spacetimes.}
\label{tab:Einstein tensor}
\begin{center}
\begin{tabular}{|p{0.4in}|p{4in}|} \hline
metric & $\frac{-1}{8\pi } (G_{\; \; \; \; \upsilon }^{\mu }) $
$[=diag(\rho ,p_{r} ,p_{t} ,p_{t} )]$ \\ \hline $1$ & $diag($$0$,
\ $\frac{-C_{1} }{8\pi r^{2} \left(r+C_{1} \right)} $, \
$\frac{C_{1} \left(2r-C_{1} \right)}{32\pi r^{2} \left(r+C_{1}
\right)^{2} } $, \ $\frac{C_{1} \left(2r-C_{1} \right)}{32\pi
r^{2} \left(r+C_{1} \right)^{2} } $$)$ \\ \hline $2$ & $diag($$0$,
\ $\frac{2C_{2} }{8\pi (1+C_{2} r^{2} )} $, \ $\frac{C_{2}
\left(2-C_{2} r^{2} \right)}{8\pi \left(1+C_{2} r^{2} \right)^{2}
} $, \ $\frac{C_{2} \left(2-C_{2} r^{2} \right)}{8\pi
\left(1+C_{2} r^{2} \right)^{2} } $$)$ \\ \hline $3$ &
$diag($$\frac{1-2r}{8\pi r^{2} } $, \ $\frac{-2C_{3} r^{3} +C_{3}
r^{2} -r^{4} +r^{3} -4C_{4} r+C_{4} }{8\pi r^{2} \left(r^{3}
+C_{3} r^{2} +C_{4} \right)} $, \ \ $\frac{-3C_{3} r^{5} -4C_{3}
^{2} r^{4} -6C_{3} C_{4} r^{2} +11C_{4} r^{3} -14C_{4} ^{2}
-2r^{6} }{32\pi r\left(r^{3} +C_{3} r^{2} +C_{4} \right)^{2} } $,
\ \ $\frac{-3C_{3} r^{5} -4C_{3} ^{2} r^{4} -6C_{3} C_{4} r^{2}
+11C_{4} r^{3} -14C_{4} ^{2} -2r^{6} }{32\pi r\left(r^{3} +C_{3}
r^{2} +C_{4} \right)^{2} } $$)$ \\ \hline $4$ &
$diag($$\frac{1-5r^{4} }{8\pi r^{2} } $, \ $\frac{-5C_{5} r^{6}
+C_{5} r^{2} -r^{10} +r^{6} -7C_{6} r^{4} +C_{6} }{8\pi r^{2}
\left(r^{6} +C_{5} r^{2} +C_{6} \right)} $, \ \ $\frac{r^{2}
(-10C_{5} ^{2} r^{4} -24C_{5} C_{6} r^{2} +8C_{6} r^{6} -17C_{6}
^{2} -2r^{12} )}{8\pi \left(r^{6} +C_{5} r^{2} +C_{6} \right)^{2}
} $, \ \ $\frac{r^{2} (-10C_{5} ^{2} r^{4} -24C_{5} C_{6} r^{2}
+8C_{6} r^{6} -17C_{6} ^{2} -2r^{12} )}{8\pi \left(r^{6} +C_{5}
r^{2} +C_{6} \right)^{2} } $$)$ \\ \hline $5$ &
$diag($$\frac{1-3r^{2} }{8\pi r^{2} } $, \ $\frac{-4C_{7} r^{3}
+C_{7} r-r^{6} +r^{4} -5C_{8} r^{2} +C_{8} }{8\pi r^{2}
\left(r^{4} +C_{7} r+C_{8} \right)} $, \ \ $\frac{4C_{7} r^{5}
-19C_{7} ^{2} r^{2} -44C_{7} C_{8} r+16C_{8} r^{4} -28C_{8} ^{2}
-4r^{8} }{32\pi \left(r^{4} +C_{7} r+C_{8} \right)^{2} } $, \ \
$\frac{4C_{7} r^{5} -19C_{7} ^{2} r^{2} -44C_{7} C_{8} r+16C_{8}
r^{4} -28C_{8} ^{2} -4r^{8} }{32\pi \left(r^{4} +C_{7} r+C_{8}
\right)^{2} } $$)$ \\ \hline $6$ & $diag($$\frac{1-3r^{2} }{8\pi
r^{2} } $, \ $\frac{-5C_{9} r^{2} +C_{9} -r^{6} +r^{4} }{8\pi
r^{2} \left(r^{4} +C_{9} \right)} $, \ \ $\frac{4C_{9} r^{4}
-7C_{9} ^{2} -r^{8} }{8\pi \left(r^{4} +C_{9} \right)^{2} } $, \
$\frac{4C_{9} r^{4} -7C_{9} ^{2} -r^{8} }{8\pi \left(r^{4} +C_{9}
\right)^{2} } $$)$ \\ \hline $7$ & $diag($$\frac{C_{11} (-2C_{10}
r+C_{11} -r^{2} )}{8\pi r^{2} \left(r^{2} +C_{11} \right)^{2} } $,
\ $\frac{C_{11} ^{2} -4C_{10} C_{11} r-3C_{11} r^{2} -2C_{10}
r^{3} -2r^{4} }{8\pi r^{2} \left(r^{2} +C_{11} \right)^{2} } $, \
\ $\frac{-2r^{5} -8C_{11} r^{3} -14C_{11} ^{2} r+C_{10} r^{4}
+2C_{10} C_{11} r^{2} -7C_{10} C_{11} ^{2} }{16\pi r\left(r^{2}
+C_{11} \right)^{3} } $, \ \ $\frac{-2r^{5} -8C_{11} r^{3}
-14C_{11} ^{2} r+C_{10} r^{4} +2C_{10} C_{11} r^{2} -7C_{10}
C_{11} ^{2} }{16\pi r\left(r^{2} +C_{11} \right)^{3} }
$$)$ \\ \hline $8$ & $diag($$\frac{-3C_{11} C_{12} r^{2} -C_{12}
r^{4} +C_{11} ^{2} -5C_{11} r^{4} +r^{4} +2C_{11} r^{2} -3r^{6}
}{8\pi r^{2} \left(r^{2} +C_{11} \right)^{2} } $, \ \
$\frac{C_{11} ^{2} -5C_{11} C_{12} r^{2} +2C_{11} r^{2} -7C_{11}
r^{4} -3C_{12} r^{4} +r^{4} -5r^{6} }{8\pi r^{2} \left(r^{2}
+C_{11} \right)^{2} } $, \ \ $\frac{-7r^{6} -20C_{11} r^{4}
-17C_{11} ^{2} r^{2} -C_{12} r^{4} -4C_{11} C_{12} r^{2} -7C_{11}
^{2} C_{12} }{8\pi \left(r^{2} +C_{11} \right)^{3} } $, \ \
$\frac{-7r^{6} -20C_{11} r^{4} -17C_{11} ^{2} r^{2} -C_{12} r^{4}
-4C_{11} C_{12} r^{2} -7C_{11} ^{2} C_{12} }{8\pi \left(r^{2}
+C_{11} \right)^{3} } $$)$ \\ \hline $9$ & $diag($$\frac{-C_{11}
C_{13} +C_{13} r^{2} +C_{11} ^{2} -C_{11} r^{2} }{8\pi r^{2}
\left(r^{2} +C_{11} \right)^{2} } $, \ $\frac{C_{11} ^{2} -2C_{11}
r^{2} -2C_{11} C_{13} -r^{4} }{8\pi r^{2} \left(r^{2} +C_{11}
\right)^{2} } $, \ \ $\frac{-r^{6} -4C_{11} r^{4} -19C_{11} ^{2}
r^{2} +C_{13} r^{4} +16C_{11} C_{13} r^{2} -C_{11} ^{2} C_{13}
}{32\pi r^{2} \left(r^{2} +C_{11} \right)^{3} } $, \ \
$\frac{-r^{6} -4C_{11} r^{4} -19C_{11} ^{2} r^{2} +C_{13} r^{4}
+16C_{11} C_{13} r^{2} -C_{11} ^{2} C_{13} }{32\pi r^{2}
\left(r^{2} +C_{11} \right)^{3} } $$)$ \\ \hline
\end{tabular}
\end{center}
\end{table}

\begin{tabular}{|p{0.4in}|p{4in}|} \hline
$10$ & $diag($$\frac{-C_{11} C_{13} +C_{13} r^{2} +C_{11} ^{2}
-C_{11} r^{2} }{8\pi r^{2} \left(r^{2} +C_{11} \right)^{2} } $, \
\ $\frac{C_{11} ^{2} +C_{11} r^{2} +C_{11} C_{13} +2r^{4} +3C_{13}
r^{2} }{8\pi r^{2} \left(r^{2} +C_{11} \right)^{2} } $, \ \
$\frac{-r^{6} +2C_{11} r^{4} -5C_{13} r^{4} -2C_{11} C_{13} r^{2}
-C_{11} ^{2} r^{2} -C_{11} ^{2} C_{13} }{8\pi r^{2} \left(r^{2}
+C_{11} \right)^{3} } $, \ \ $\frac{-r^{6} +2C_{11} r^{4} -5C_{13}
r^{4} -2C_{11} C_{13} r^{2} -C_{11} ^{2} r^{2} -C_{11}
^{2} C_{13} }{8\pi r^{2} \left(r^{2} +C_{11} \right)^{3} } $$)$ \\
\hline $11$ & $diag($$\frac{-2C_{11} r+C_{14} r^{2} +r^{4}
+2C_{11} r^{2} +C_{11} ^{2} -C_{11} C_{14} }{8\pi r^{2}
\left(r^{2} +C_{11} \right)^{2} } $, \   \  $\frac{-2C_{11}
r+C_{14} r^{2} +r^{4} +2C_{11} r^{2} +C_{11} ^{2} -C_{11} C_{14}
}{8\pi r^{2} \left(r^{2} +C_{11} \right)^{2} } $, \ \
$\frac{3C_{11} C_{14} r-C_{14} r^{3} +3C_{11} r^{2} -C_{11} ^{2}
}{8\pi r\left(r^{2} +C_{11} \right)^{3} } $, \ $\frac{3C_{11}
C_{14} r-C_{14} r^{3} +3C_{11} r^{2} -C_{11} ^{2} }{8\pi
r\left(r^{2} +C_{11} \right)^{3} } $$)$ \\ \hline $12$ &
$diag($$\frac{-3r^{6} -5C_{11} r^{4} +C_{15} r^{2} +r^{4} +2C_{11}
r^{2} +C_{11} ^{2} -C_{11} C_{15} }{8\pi r^{2} \left(r^{2} +C_{11}
\right)^{2} } $, \ \ $\frac{-3r^{6} -5C_{11} r^{4} +C_{15} r^{2}
+r^{4} +2C_{11} r^{2} +C_{11} ^{2} -C_{11} C_{15} }{8\pi r^{2}
\left(r^{2} +C_{11} \right)^{2} } \  $,\  $\frac{-C_{15} r^{2}
+3C_{11} C_{15} -9C_{11} r^{4} -10C_{11} ^{2} r^{2} -3r^{6} }{8\pi
\left(r^{2} +C_{11} \right)^{3} } $, \ \  $\frac{-C_{15} r^{2}
+3C_{11} C_{15} -9C_{11} r^{4} -10C_{11} ^{2} r^{2} -3r^{6} }{8\pi
\left(r^{2} +C_{11} \right)^{3} } $$)$ \\ \hline $13$ &
$diag($$\frac{r^{2} +1}{8\pi r^{4} } $, \ $\frac{C_{10} r^{5}
+r^{6} +r^{4} +C_{8} r^{2} -3C_{8} }{8\pi r^{4} \left(r^{4}
+C_{10} r^{3} +C_{8} \right)} $, \ \  $\frac{-4r^{8} +C_{10} ^{2}
r^{6} +32C_{8} C_{10} r^{3} +4C_{8} ^{2} +48C_{8} r^{4} }{32\pi
r^{4} \left(r^{4} +C_{10} r^{3} +C_{8} \right)^{2} } $, \ \
$\frac{-4r^{8} +C_{10} ^{2} r^{6} +32C_{8} C_{10} r^{3} +4C_{8}
^{2} +48C_{8} r^{4} }{32\pi r^{4} \left(r^{4} +C_{10} r^{3} +C_{8}
\right)^{2} } $$)$
\\ \hline $14$ & $diag($$\frac{r^{2} +1}{8\pi r^{4} } $, \
$\frac{C_{12} r^{6} +C_{12} r^{4} +r^{8} +3r^{6} +C_{6} r^{2}
-3C_{6} }{8\pi r^{4} \left(r^{6} +C_{12} r^{4} +C_{6} \right)} $,
\ \ $\frac{-5r^{12} -3C_{12} r^{10} +23C_{6} r^{6} -C_{12} ^{2}
r^{8} +C_{6} ^{2} +12C_{6} C_{12} r^{4} }{8\pi r^{4} \left(r^{6}
+C_{12} r^{4} +C_{6} \right)^{2} } $, \ \ $\frac{-5r^{12} -3C_{12}
r^{10} +23C_{6} r^{6} -C_{12} ^{2} r^{8} +C_{6} ^{2} +12C_{6}
C_{12} r^{4} }{8\pi r^{4} \left(r^{6} +C_{12} r^{4} +C_{6}
\right)^{2} } $$)$ \\ \hline $15$ & $diag($$\frac{1}{8\pi r^{2} }
$, \ $\frac{C_{13} r^{2} -2C_{13} r+r^{4} +C_{7} r-3C_{7} }{8\pi
r^{3} \left(r^{3} +C_{13} r+C_{7} \right)} $, \ \ $\frac{14C_{13}
r^{3} +2C_{13} ^{2} r+5C_{7} C_{13} +27C_{7} r^{2} }{32\pi r^{2}
\left(r^{3} +C_{13} r+C_{7} \right)^{2} } $, \ $\frac{14C_{13}
r^{3} +2C_{13} ^{2} r+5C_{7} C_{13} +27C_{7} r^{2}
}{32\pi r^{2} \left(r^{3} +C_{13} r+C_{7} \right)^{2} } $$)$ \\
\hline $16$ & $diag($$\frac{1-3r^{2} }{8\pi r^{2} } $, \
$\frac{-5C_{16} r^{2} +C_{16} -3r^{4} +r^{2} }{8\pi r^{2}
\left(r^{2} +C_{16} \right)} $, \ \  $\frac{-3r^{4} -7C_{16} r^{2}
-7C_{16} ^{2} }{8\pi \left(r^{2} +C_{16} \right)^{2} } $, \
$\frac{-3r^{4} -7C_{16} r^{2} -7C_{16} ^{2} }{8\pi \left(r^{2}
+C_{16} \right)^{2} } $$)$ \\ \hline $17$ & $diag($$0$, \
$\frac{-r-2C_{17} }{8\pi r^{2} \left(r+C_{17} \right)} $, \
$\frac{-r^{2} -2C_{17} r-4C_{17} ^{2} }{32\pi r^{2} \left(r+C_{17}
\right)^{2} } $, \ $\frac{-r^{2} -2C_{17} r-4C_{17} ^{2} }{32\pi
r^{2} \left(r+C_{17} \right)^{2} } $$)$ \\ \hline $18$ &
$diag($$0$, \ $\frac{2(r^{4} -C_{18} )}{8\pi r^{2} \left(r^{4}
+C_{18} \right)} $, \ $\frac{-r^{8} -C_{18} ^{2} +10C_{18} r^{4}
}{8\pi r^{2} \left(r^{4} +C_{18} \right)^{2} } $, \ $\frac{-r^{8}
-C_{18} ^{2} +10C_{18} r^{4} }{8\pi r^{2} \left(r^{4} +C_{18}
\right)^{2} } $$)$ \\ \hline
 \end{tabular}\\

\begin{table}
\caption{The Ricci scalar for all the 18 classes of spacetimes.}
\label{tab:Ricci scalar}
\begin{center}
\begin{tabular}{|p{0.4in}|p{4in}|} \hline
$metric$ & \textit{$R$} \\ \hline
$1$ & $[ - 3{C_1}^2]/[2{r^2}{\left( {r + {C_1}} \right)^2}]$\\ \hline
$2$ & $[ - 6{C_2}]/[{\left( {1 + {C_2}{r^2}} \right)^2}]$ \\ \hline
$3$ & $\begin{array}{l}[ - (8{r^7} + (17{C_3} - 4){r^6} + 4{C_3}(3{C_3} - 2){r^5} + (7{C_4} - 4{C_3}^2){r^4} + \\2{C_4}(13{C_3} - 4){r^3} - 8{C_3}{C_4}{r^2} + 26{C_4}^2r - 4{C_4}^2)]\\/[2{r^2}{\left( {{r^3} + {C_3}{r^2} + {C_4}} \right)^2}]\end{array}$ \\ \hline
$4$ & $\begin{array}{l}[ - 2(5{r^{16}} + (8{C_5} - 1){r^{12}} + {C_6}{r^{10}} + {C_5}(15{C_5} - 2){r^8} \\+ {C_6}(35{C_5} - 2){r^6} + (23{C_6}^2 - {C_5}^2){r^4} - 2{C_5}{C_6}{r^2} - {C_6}^2)]\\/[{r^2}{\left( {{r^6} + {C_5}{r^2} + {C_6}} \right)^2}]\end{array}$ \\ \hline
$5$ & $\begin{array}{l}[ - (12{r^{10}} - 4{r^8} + 18{C_7}{r^7} + 8{C_8}{r^6} - 8{C_7}{r^5} + (33{C_7}^2 - 8{C_8}){r^4} \\+ 74{C_7}{C_8}{r^3} + 4(11{C_8}^2 - {C_7}^2){r^2} - 8{C_7}{C_8}r - 4{C_8}^2)]\\/[2{r^2}{\left( {{r^4} + {C_7}r + {C_8}} \right)^2}]\end{array}$ \\ \hline
$6$ & $[ - 2(3{r^{10}} - {r^8} + 2{C_9}{r^6} - 2{C_9}{r^4} + 11{C_9}^2{r^2} - {C_9}^2)]/[{r^2}{\left( {{r^4} + {C_9}} \right)^2}]$ \\ \hline
$7$ & $\begin{array}{l}[ - 4{r^6} + {C_{10}}{r^5} + 14{C_{11}}{r^4} + 6{C_{10}}{C_{11}}{r^3} + 16{C_{11}}^2{r^2} + \\13{C_{10}}{C_{11}}^2r - 2{C_{11}}^2]/[{r^2}{\left( {{r^2} + {C_{11}}} \right)^3}]\end{array}$ \\ \hline
$8$ & $\begin{array}{l}[2( - 11{r^8} + \left( {1 - 30{C_{11}} - 3{C_{12}}} \right){r^6} +\\ {C_{11}}(3 - 23{C_{11}} - 10{C_{12}}){r^4} + {C_{11}}^2(3 - 11{C_{12}}){r^2} + {C_{11}}^3)]\\/[{r^2}{\left( {{r^2} + {C_{11}}} \right)^3}]\end{array}$ \\ \hline
$9$ & $\begin{array}{l}[ - 3{r^6} + 3\left( {{C_{13}} - 4{C_{11}}} \right){r^4} + 3{C_{11}}\left( {4{C_{13}} - 7{C_{11}}} \right){r^2} + \\{C_{11}}^2(4{C_{11}} - 7{C_{13}})]/[2{r^2}{\left( {{r^2} + {C_{11}}} \right)^3}]\end{array}$ \\ \hline
$10$ & $[2\left( {{C_{11}} - {C_{13}}} \right)(3{r^4} + {C_{11}}^2)]/[{r^2}{\left( {{r^2} + {C_{11}}} \right)^3}]$ \\ \hline
$11$ & $\begin{array}{l}[2({r^6} + 3{C_{11}}{r^4} + {C_{11}}{r^3} + 3{C_{11}}({C_{11}} + {C_{14}}){r^2}\\ - 3{C_{11}}^2r + {C_{11}}^2({C_{11}} - {C_{14}}))]/[{r^2}{\left( {{r^2} + {C_{11}}} \right)^3}]\end{array}$ \\ \hline
$12$ & $\begin{array}{l}[2( - 6{r^8} + \left( {1 - 17{C_{11}}} \right){r^6} + 3{C_{11}}(1 - 5{C_{11}}){r^4} + \\3{C_{11}}({C_{11}} + {C_{15}}){r^2} + {C_{11}}^2({C_{11}} - {C_{15}}))]/[{r^2}{\left( {{r^2} + {C_{11}}} \right)^3}]\end{array}$ \\ \hline
$13$ & $\begin{array}{l}[4{r^8} + 8{C_{10}}{r^7} + 4{C_{10}}^2{r^6} + 6{C_{10}}{r^5} + (3{C_{10}}^2 + 8{C_8}){r^4}\\ + 8{C_8}{C_{10}}{r^3} + 48{C_8}{r^2} + 30{C_8}{C_{10}}r + 4{C_8}^2]\\/[2{r^2}{\left( {{r^4} + {C_{10}}{r^3} + {C_8}} \right)^2}]\end{array}$ \\ \hline
$14$ & $\begin{array}{l}[2({r^{12}} + (2{C_{12}} - 3){r^{10}} + {C_{12}}^2{r^8} + 2{C_6}{r^6} + 2{C_6}({C_{12}} + 12){r^4}\\ + 12{C_6}{C_{12}}{r^2} + {C_6}^2)]/[{r^2}{\left( {{r^6} + {C_{12}}{r^4} + {C_6}} \right)^2}]\end{array}$ \\ \hline
$15$ & $\begin{array}{l}[4{r^7} + 8{C_{13}}{r^5} + 2\left( {5{C_{13}} + 4{C_7}} \right){r^4} + \left( {4{C_{13}}^2 + 21{C_7}} \right){r^3} \\+ 2{C_{13}}(4{C_7} -{C_{13}}){r^2} + {C_7}(4{C_7} - 5{C_{13}})r - 6{C_7}^2]\\/[2{r^3}{\left( {{r^3} + {C_{13}}r + {C_7}} \right)^2}]\end{array}$ \\ \hline
$16$ & $[ - 2(6{r^6} + (14{C_{16}} - 1){r^4} + {C_{16}}(11{C_{16}} - 2){r^2} - {C_{16}}^2)]/[{r^2}{\left( {{r^2} + {C_{16}}} \right)^2}]$ \\ \hline
$17$ & $[ - (3{r^2} + 8{C_{17}}r + 8{C_{17}}^2)]/[2{r^2}{\left( {r + {C_{17}}} \right)^2}]$ \\ \hline
$18$ & $[4{C_{18}}(5{r^4} - {C_{18}})]/[{r^2}{\left( {{r^4} + {C_{18}}} \right)^2}]$ \\ \hline
\end{tabular}
\end{center}
\end{table}

\begin{table}
\caption{Some properties of the spacetimes (the sign $*$ means that the equations are higher than order 3, and no conclusion is reached).}
\label{tab:properties}
\begin{center}
\begin{tabular}{|p{0.4in}|p{0.8in}|p{1.5in}|p{0.4in}|p{1.2in}|} \hline
$metric$ & \textit{$M$ \  $(Tolman \   mass)$} & \textit{ $r_{m} \
(singularity \ coord.)$} & \textit{ $\mathop{\lim }\limits_{r\to
\infty } R$} & $Energy$ $Conditions$\  ($WEC$ $\& $ $SEC$)
\\ \hline $1$ & $-\frac{1}{\; 2} C_{1} $ & $-C_{1} $
\    $\; (if\; \; C_{1} <0)$ & $0$ & $violated$ \\ \hline $2$ &
$\frac{1}{\sqrt{C_{2} } } $\  $(if\; \; C_{2} >0)$ &
$\frac{1}{\sqrt{-C_{2} } } $   \   $(if\; \; C_{2} <0)$ & $0$ &
$r<\sqrt{\frac{2}{\; C_{2} } } \; \; \; $\  $(C_{2}
>0)$\  $WEC$ $\&$ $SEC$ $satisfied$ \\ \hline $3$ & $\infty
$  \  $(if\; \; C_{3} <0)$\  $-\infty $  \  $(if\; \; C_{3}
>0)$ & $0$,\  $\frac{2C_{{\rm 3}} ^{2} }{3\sqrt[{3}]{C_{{\rm
19}} } } +\frac{\sqrt[{3}]{C_{{\rm 19}} } }{6} -\frac{C_{{\rm 3}}
}{3} $ \  $\begin{array}{l} {(C_{{\rm 19}} =-8C_{{\rm 3}} ^{3} +}
\\ {12\sqrt{81C_{{\rm 4}} ^{2} +12C_{{\rm 4}} C_{{\rm 3}} ^{3} }
-} \\ {108C_{{\rm 4}} )}
\end{array}$ & $0$ & $*$ \\ \hline $4$ & $0$ & $0$,\
$\left(\frac{C_{20} ^{\frac{2}{3} } -12C_{{\rm 5}} }{\left(6C_{20}
\right)^{\frac{1}{3} } } \right)^{\frac{1}{2} } $\
$\begin{array}{l} {(C_{20} =-108C_{{\rm 6}} +} \\
{12\left(12C_{{\rm 5}} ^{3} +81C_{{\rm 6}} ^{2}
\right)^{\frac{1}{2} } )} \end{array}$ & $-\infty $ & $*$ \\
\hline $5$ & $0$ & $0$,\    $*$ & $const.$ & $*$ \\ \hline $6$ &
$0$ & $0$, \  $\sqrt[{4}]{-C_{9} } $\  $(if\; \; C_{9} <0)$ &
$const.$ & $*$ \\ \hline $7$ & $-\infty $ & $0$,\  $-C_{10} $\
$(if\; \; C_{10} <0)$,\
$\sqrt{-C_{11} } $\  $(if\; \; C_{11} <0)$ & $0$ & $*$ \\
\hline $8$ & $-\infty $ & $0$,\  $\sqrt{-C_{12} } $\  $(if\; \;
C_{12} <0)$,\   $\sqrt{-C_{11} } $\   $(if\; \; C_{11} <0)$ &
$const.$ & \  \  $*$ \\ \hline $9$ & $\infty $ & $\sqrt{-C_{11} }
$\  $(if\; \; C_{11} <0)$,\
$\sqrt{-C_{13} } $\  $(if\; \; C_{13} <0)$ & $0$ & $*$ \\
\hline
\end{tabular}
\end{center}
\end{table}
\begin{table}
\begin{center}
\begin{tabular}{|p{0.4in}|p{0.8in}|p{1.5in}|p{0.4in}|p{1.2in}|} \hline
$10$ & $\infty $\  $(if\; \; C_{13} >C_{11} )$\  $-\infty $\
$(if\; \; C_{13} <C_{11} )$ & $0$,\  $\sqrt{-C_{11} } \; $\ $(if\;
\; C_{11} <0)$,\
$\sqrt{-C_{13} } $\  $(if\; \; C_{13} <0)$ & $0$ & $*$ \\
\hline $11$ & $-\infty $ & $\sqrt{-C_{11} } $ \  $(if\; \; C_{11}
<0)$,\  $-C_{14} $\  $(if\; \; C_{14} <0)$ & $0$ & $*$
\\ \hline $12$ & $-\infty $ & $\sqrt{-C_{11} } $\  $(if\; \;
C_{11} <0)$,\  $\sqrt[{4}]{-C_{15} } $\  $(if\; \; C_{15} <0)$\ &
$const.$ & $*$ \\ \hline $13$ & $0$ & $*$ & $0$ & $*$ \\ \hline
$14$ & $0$ & $\left(\frac{C_{{\rm 21}} ^{\frac{2}{3} } +4C_{{\rm
12}} ^{2} -2C_{{\rm 12}} C_{{\rm 21}} ^{\frac{1}{3} }
}{6^{\frac{1}{2} } C_{{\rm 21}} ^{\frac{1}{3} } }
\right)^{\frac{1}{2} } $\  $\begin{array}{l} {(C_{{\rm 21}}
=-108C_{{\rm 6}} -8C_{{\rm 12}} ^{3} } \\ {+12\left(81C_{{\rm 6}}
^{2} +12C_{{\rm 6}} C_{{\rm 12}} ^{3} \right)^{\frac{1}{2} } )}
\end{array}$  & $0$ & $*$ \\ \hline $15$ & $\frac{1}{2} $ &
$\frac{C_{{\rm 22}} ^{\frac{2}{3} } -12C_{{\rm 13}} }{6C_{{\rm
22}} ^{\frac{1}{3} } } $\  $\begin{array}{l} {(C_{{\rm 22}}
=-108C_{{\rm 7}} +} \\ {12\left(12C_{{\rm 13}} ^{3} +81C_{{\rm 7}}
^{2} \right)^{\frac{1}{2} } )} \end{array}$ & $0$ & $*$ \\ \hline
$16$ & $-\infty $ &    $0$,\  $\sqrt{-C_{16} } $\  $(if\; \;
C_{16} <0)$\   & $const.$ & $*$ \\ \hline $17$ & $-\infty $
& $-C_{17} $\  $(if\; \; C_{17} <0)$ & $0$ & $violated$ \\
\hline $18$ & $1$ & $\sqrt[{4}]{-C_{18} } $\  $(if\; \; C_{18}
<0)$ & $0$  & $r\ge (5+2\sqrt{6} )^{\frac{1}{4} }
\sqrt[{4}]{C_{18} } $  \  $\; C_{18} \ge 0$\  $WEC$ $satisfied$,\
$r\ge (\frac{1}{5^{\frac{1}{4} } } )\sqrt[{4}]{C_{18} } $ \   $\;
C_{18} \ge 0$\  $SEC$ $satisfied$\  ($WEC$ $\&$ $SEC$ $satisfied$)
\\ \hline
\end{tabular}
\end{center}
\end{table}

\section{Relevance to Bertrand spacetimes}
\indent In this section, we  introduce the Bertrand spacetimes, and discuss their relevance to some of the spacetimes of previous sections.\\
\indent  A spacetime is called a ``Bertrand spacetime'' if it is a
spherically symmetric and static spacetime, and there is a
circular trajectory through each point and the following
inequality is satisfied
\begin{equation} \label{59}
0<\frac{rB'(r)}{B(r)} <1.
\end{equation}
Also, it is required that  any initial condition for the geodesic
equation which is sufficiently close to a circular trajectory
gives a periodic trajectory \cite{Perlick1992}.\\
Bertrand spacetimes are given by the following metrics, which are
called types \textit{I} and \textit{II}$\pm $
\cite{Perlick1992,Ballesteros2008}
\begin{equation} \label{60}
ds^{2} =-\frac{dt^{2} }{G+\sqrt{r^{-2} +K} } +\frac{dr^{2} }{\beta ^{2} \left(1+Kr^{2} \right)} +r^{2} d\theta ^{2} +r^{2} sin^{2} \theta d\varphi ^{2} \ \ \        ({\rm Type\; }I\; ),
\end{equation}
and
\[ds^{2} =-\frac{dt^{2} }{G\mp r^{2} [1-Dr^{2} \pm \sqrt{(1-Dr^{2} )^{2} -Kr^{4} } ]^{-1} } \\ \]
\begin{equation} \label{61}
+\frac{2(1-Dr^{2} \pm \sqrt{(1-Dr^{2} )^{2} -Kr^{4} } )}{\beta ^{2} ((1-Dr^{2} )^{2} -Kr^{4} )} dr^{2} + r^{2} d\theta ^{2} +r^{2} sin^{2} \theta d\varphi ^{2} \ \ \ ({\rm Type\; }II\pm ),
\end{equation}
the parameters \textit{G}, \textit{K} and \textit{D} are real constants,
and $\beta $\textbf{\textit{ }}is a positive rational number.
Conversely, any metric of this form determines a Bertrand spacetime.\\
\indent According to \cite{Ballesteros2008}, there are several
relevant specific cases of Bertrand spaces: (i) Three classical
Riemannian spaces of constant curvature, (ii) Darboux spaces of
type III, and (iii) Iwai--Katayama \cite{Iwai1995} spaces. The
classical Riemannian spaces are proven to belong to the Bertrand
family \textit{I}  \ref{60}, and \textit{II}$\pm $\textit{
}\ref{61}, respectively under the identifications
\cite{Ballesteros2008}
\begin{equation} \label{62}
\beta =1,  \ \ \     K=-\; \alpha ,    \ \ \   ({\rm Type\; }I\; ),
\end{equation}
\begin{equation} \label{63}
\beta =2,  \ \ \     K=0,    \ \ \      D=\alpha ,     \ \ \    ({\rm Type\; }II+),
\end{equation}
where $\alpha $ is a constant. And the Darboux spaces and the
Iwai--Katayama spaces are proven to belong to the Bertrand family
\textit{II}$\pm $ \ref{61}, respectively, under the
identifications \cite{Ballesteros2008}
\begin{equation} \label{64}
\beta =2,    \ \ \    K=D^{2} ,   \ \ \    D=-\frac{2}{l^{2} } ,   \ \ \    ({\rm Type\; }II\pm ),
\end{equation}
\begin{equation} \label{65}
\beta =\frac{1}{\upsilon } ,  \ \ \     K=D^{2} ,   \ \ \   D=-\frac{2b}{a^{2} } ,    \ \ \   ({\rm Type\; }II\pm ),
\end{equation}
where $l$ is an arbitary constant, $\upsilon $ is a rational number, and $a$ and $b$ are two real constants.\\
\indent It is easy to show that the metric \ref{40} belongs to the
Bertrand family  $I$ \ref{60} under the identifications
\begin{equation} \label{66}
\beta =1,    \ \ \     K=0,      \ \ \   ({\rm Type\; }I\; ),
\end{equation}
($G=\frac{1}{C_1}$). And the metric \ref{42} belongs to the
Bertrand family  \textit{II}$+$ \ref{61} under the identifications
\begin{equation} \label{67}
\beta =2,    \ \ \     K=0,     \ \ \   D=0,      \ \ \    ({\rm Type\; }II+),
\end{equation}
($G=\frac{-1}{2C_2}$). We therefore see that \ref{40} and \ref{42} are particular cases of the Bertrand spacetimes. We could not identify other metrics of table \ref{tab:all spacetimes} with Bertrand spacetimes.\\

\section{Concluding remarks}
\indent We derived metrics for curved spacetimes which lead to closed bound projected orbits by comparing the relativistic differential equation of orbits with the corresponding classical differential equation. We can name this method as the form-compatibility method.\\
\indent Physical and geometrical properties of these peculiar spacetimes were derived and discussed and it was shown that two of these spacetimes may be perfect fluid solutions of the Einstein's field equations (spacetimes 2 and 18 in table \ref{tab:all spacetimes}). It was also shown that two of these spacetimes are particular cases of the Bertrand spacetimes (spacetimes 1 and 2 in table \ref{tab:all spacetimes}).\\
\indent In classical mechanics, the existence of closed orbits guides us toward discovering further symmetries of the dynamical system (as established for the Kepler and harmonic potentials via the existence of the Laplace-Runge-Lenz vector). The Laplace-Runge-Lenz vector is in the direction of the radius vector to the perihelion point on the orbit. Conservation of this vector means that the orientation of the orbit in space is fixed and the orbit stays closed.\\
\indent In general relativity, the existence of projected closed orbits guides us toward discovering further symmetries of the spacetime. The geodesic equation of a Bertrand spacetime admits an additional constant of motion related to a non-redundant time-independent second rank Killing tensor field if and only if $\beta $\textbf{\textit{ }}is equal either to 1\textbf{ }or to 2 \cite{Perlick1992}. Two of our spacetimes, have this condition (spacetimes 1 and 2 in table \ref{tab:all spacetimes}). All spherically symmetric and static spacetimes which admit non-redundant time independent second rank Killing tensor fields are listed in reference \cite{Hauser1974}. The corresponding constants of motion are explicitly given in table 2 of this reference. The whole problem of hidden symmetries in general relativity is remains a subject of much interest.\\

\end{document}